# Jordan Journal of Physics

## ARTICLE

## Structural and Mössbauer Studies of Ball Milled Co-Zn Y-type Hexaferrites

M. A. Awawdeh[a], R.N. Al-Bashaireh[a], A-F. Lehlooh[a] and S.H. Mahmood[b]

[a] *Department of Physics, Yarmouk University, Irbid, Jordan.*
[b] *Department of Physics, The University of Jordan, Amman, 119-42, Jordan.*

**Abstract:** Y-type hexaferrite samples $Ba_2Zn_xCo_{2-x}Fe_{12}O_{22}$ with $x$ = 0, 1 and 2 were prepared from precursors synthesized by the citrate sol-gel auto combustion method. The powders were sintered at 1100° C for four hours. X-ray diffraction, scanning electron microscopy and Mössbauer spectroscopy were used to investigate the effect of ball milling on the structure, microstructure and on the hyperfine parameters. X-ray diffraction revealed the presence of a single Y-type hexaferrite in all sintered samples. This phase persisted in samples milled for periods up to four hours. Scanning electron microscopy images indicated that the particles changed shape, and their sizes reduced by a factor of 10 for milling times of four hours. Room temperature Mössbauer spectra of the hexaferrites elucidated the distribution of the various cations at the various sites of the hexaferrite lattice. The hyperfine field ($B_{hf}$) associated with each component was found to decrease almost exponentially with increasing milling time. The reduction in hyperfine field was associated with the weakening of the superexchange interactions between the spin-up and spin-down sublattices as a consequence of replacing $Co^{2+}$ magnetic ions by non-magnetic $Zn^{2+}$ ions and with the microstructure in the milled samples.

## Introduction

The fabrication and characterization of different types of hexagonal ferrites and their modification by selected types of cationic substitutions had attracted the interest of a large community of scientists and technologists in the past few decades. Interest in these materials was stimulated by their potential for a wide range of technological and industrial applications, specifically, in permanent magnets, high density magnetic recording, high frequency telecommunication devices and multi-layer electronic components [1 - 6].

Hexaferrites can be classified as M-, W-, X-, Y-, Z- and U-type ferrites depending on their chemical formula and crystal structure [4, 6, 7].

The Y-type hexaferrite (also called ferroxplana), with the chemical formula $Ba_2Me_2Fe_{12}O_{22}$, has a unit cell TST'S'T"S", where the primes indicate rotation by 120° about the c-axis. The Ba ions in the structure could be replaced completely or partially by other cations such as Sr or Pb, and Me are divalent ions. This structure has an easy plane perpendicular to the *c*-axis (thus the name ferroxplana) with space group ($R\bar{3}m$) and lattice parameters: *a* = 5.89 Å and *c* = 43.56 Å [4]. The metallic cations (Me) and iron ions are distributed in six crystallographic sites: two tetrahedral sites ($6c_{IV}$ and $6c_{IV}$*) and four octahedral sites ($3a_{VI}$, $18h_{VI}$, $6c_{VI}$ and $3b_{VI}$) as indicated in Table 1 [2, 8, 9].

Corresponding Author:  M. A. Awawdeh      **Email**: amufeed@yu.edu.jo.

TABLE 1. Metallic sublattices of Y structures and their properties

| Sublattices | Coordination | Block | Number of ions per unit cell | Spin |
|---|---|---|---|---|
| $6c_{IV}$ | Tetra | S | 6 | Down |
| $3a_{VI}$ | Octa | S | 3 | Up |
| $18h_{VI}$ | Octa | S-T | 18 | Up |
| $6c_{VI}$ | Octa | T | 6 | Down |
| $6c_{IV*}$ | Tetra | T | 6 | Down |
| $3b_{VI}$ | Octa | T | 3 | Up |

Recent Mössbauer and magnetic studies on BaM and SrM hexaferrites have demonstrated the importance of the type and concentration of metal substitutions for Fe on the magnetic anisotropy and saturation magnetization of the prepared systems [10-15]. However, little had been reported in the literature concerning the preferential site occupation of the metal cations in Y-type hexaferrites, with conflicting reports in some cases [16-18].

It is well known that the grain size of a magnetic powder has a significant influence on the magnetic properties of the substance. Hexaferrites prepared by the solid state reaction method have typical grain size exceeding the critical single domain size of about 0.5 μm. In such grains, magnetization processes are influenced by domain wall motion, which leads to significant reduction in the coercivity of the powder sample. However, if the grain size is reduced below this value, the grains become single-domain and the coercivity improves significantly. Further reduction of the grain size below 100 nm leads to the development of a superparamagnetic powder, with significantly reduced coercivity and remanance magnetization which approach zero at a grain size of 10 – 20 nm. Although such systems are of no importance to applications requiring high coercivity and remanence, they are still of great importance for magnetic switching applications [4]. Further, magnetic nano-particles have promising biomedical applications such as MRI contrast agents, hyperthermia, target drug delivery and cellular imaging [4]. The range of grain size of the magnetic material is therefore of critical importance to the type of the required application.

The present work is concerned with the synthesis of Co-Zn Y-type hexaferrite powders with reduced particle size. The samples were prepared using sol-gel method and appropriate sintering. To reduce the particle size, the prepared compounds were ball milled for different milling times, then characterized using X-ray diffraction and Mössbauer spectroscopy to have an understanding of the effect of milling on the hyperfine parameters of the system.

**Experimental Procedures**

A series of samples $Ba_2Zn_xCo_{2-x}Fe_{12}O_{22}$ with x = 0.0, 1.0 and 2.0 were prepared using the citrate sol-gel auto combustion method explained in a previous publication [17]. Appropriate amounts of high purity $Fe(NO_3)_3 \_ 9H_2O$, $Ba(NO_3)_2$, $Zn(NO_3)_2 \_ 6H_2O$ and $Co(NO_3)_2$ were dissolved in distilled water, and an aqueous solution of citric acid was then added at 80 °C to chelate $Ba^{+2}$ and $Fe^{+3}$ ions in the solution. The pH of the solution was maintained at 7 by dropwise addition of ammonia. The solution was then slowly heated until the water evaporated resulting in a viscous gel, which has subsequently undergone a self-propagating combustion resulting in a loose powder. The resulting powder was preheated at 450 °C for 1 h and then sintered in air at 1100 °C for 4 h using a heating rate of 10 °C /min.

Planetary ball mill is a useful tool for reducing the particle size of a powder to an order of 100 nm, and thus obtaining single-domain particles [19]. The samples were ball-milled at different times using a Fritch type planetary ball mill. The rotating speed of the mill was set to 400 rpm. The bowls and balls of the mill are made of Tungsten Carbide and the powder-ball ratio was 1:20.

Mössbauer samples were prepared as a thin circular layer of the system powder pressed gently between two Teflon disks with a diameter of 2 cm, which is small compared with the distance between the source and the sample in order to avoid angular broadening of the spectrum. Mössbauer spectra were collected using a standard constant acceleration Mössbauer spectrometer over 1024 channels. The Mössbauer spectra were analyzed using standard fitting routines to obtain the hyperfine parameters.

The samples for the structural studies were prepared by sprinkling the powders on a microscope glass that has no structural peaks. The crystalline structure of each sample was investigated using (θ-2θ) X-ray diffractometer (Philips PW1729).

The morphology of the prepared samples was examined by scanning electron microscope (SEM FEI Quanta 200) equipped with energy dispersive X-ray spectroscopy (EDAX) facility for elemental analysis.

## Results and Discussion

### XRD Measurements

The XRD patterns were obtained for the samples milled at different times. It is obvious that milling for relatively short time periods (≤ 4 h) does not change the patterns appreciably as sown in Fig.1 – Fig. 3 for the samples with different $x$ values. All patterns show single-phase, and the pattern for the sample with $x = 0$ is consistent with the standard pattern for $Co_2Y$ hexaferrite (JCPDS 00-044-0206), while the pattern for the sample with $x = 2$ is consistent with the standard pattern for $Zn_2Y$ hexaferrite phase (JCPDS: 00-044-0207). However, the XRD peaks broaden, and the intensities reduce with increasing milling time, indicating the reduction in crystallite size.

The average crystallite sizes for the samples were calculated using Scherrer formula:

$$D = k\lambda/\beta\cos\theta$$

where; D is the crystallite size, k is the Scherrer constant (0.94), λ is the wavelength of radiation (1.542 Å), β is the peak width at half maximum measured in radians and θ is the peak position. Table 2 shows the crystallite sizes for samples milled at 4 h to be compared with those for the un-milled samples.

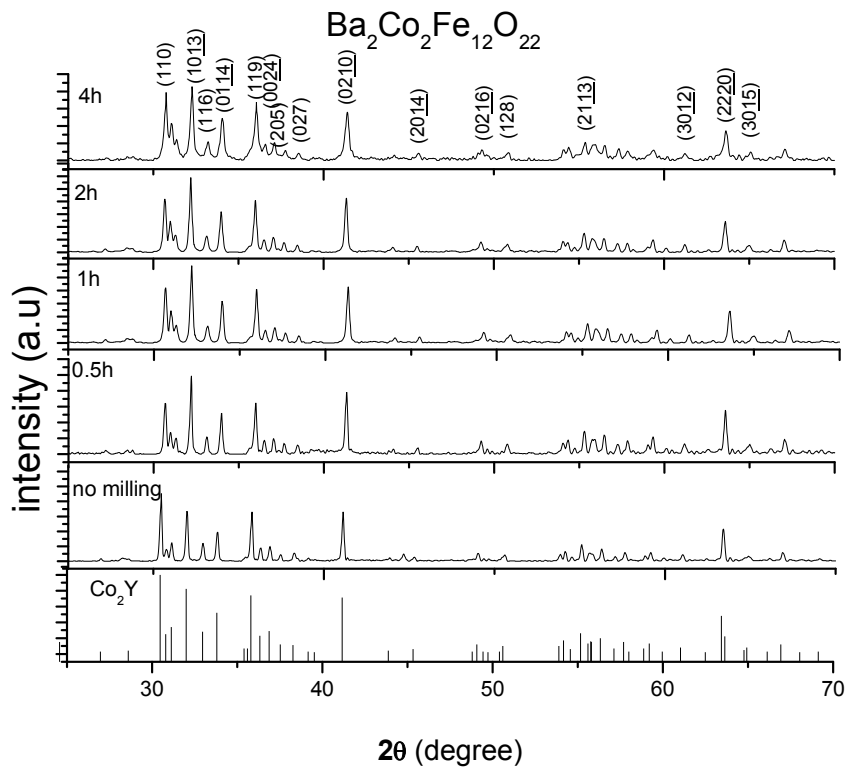

FIG. 1. XRD pattern for $Ba_2Co_2Fe_{12}O_{22}$ with different milling times

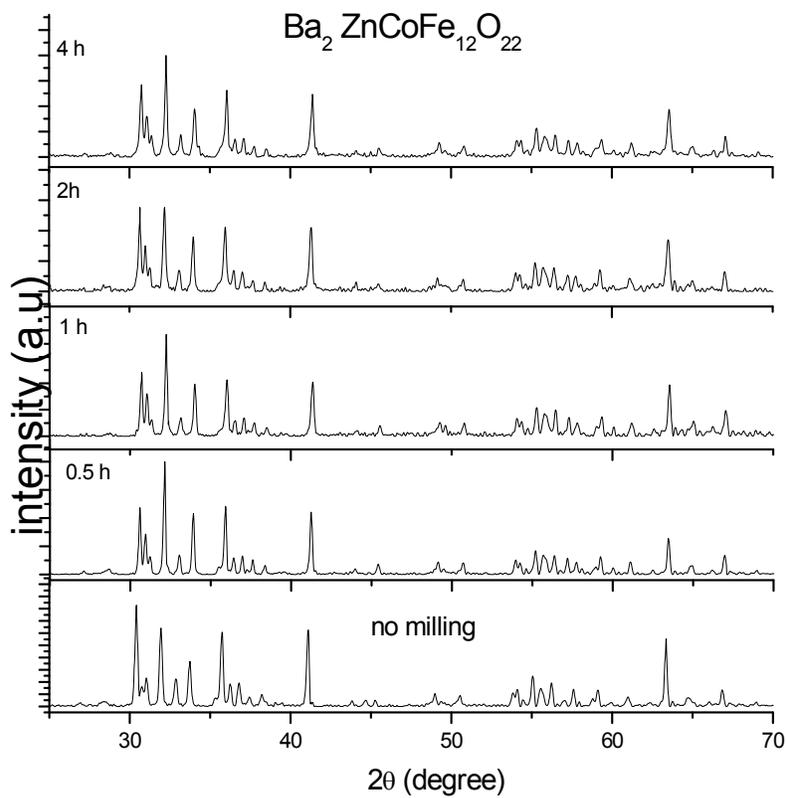

FIG. 2. XRD pattern for $Ba_2ZnCoFe_{12}O_{22}$ with different milling time

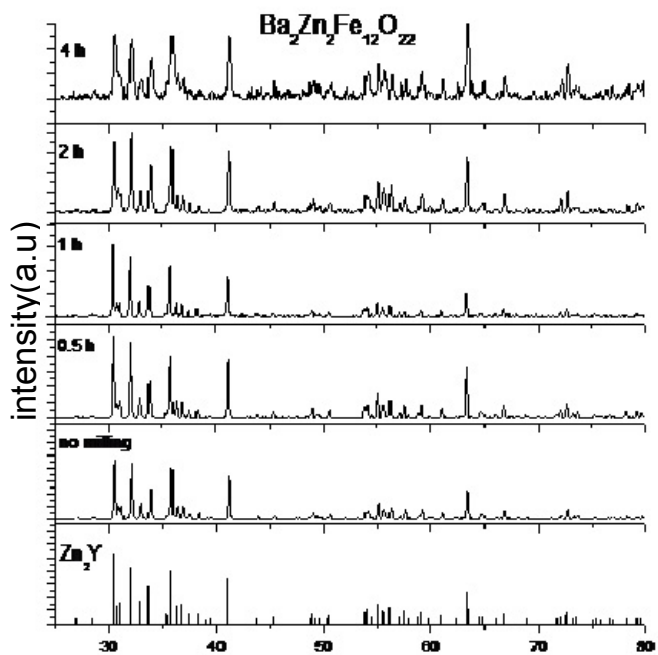

FIG. 3. XRD pattern for $Ba_2Zn_2Fe_{12}O_{22}$ with different milling times

TABLE 2. Average crystallite size for $Ba_2Zn_xCo_{2-x}Fe_{12}O_{22}$ powders milled for 4 hours

| Zn concentration (x) | Crystallite size before milling (nm) | Crystallite size after 4 h milling (nm) |
|---|---|---|
| 0.0 | 61 ± 5 | 32 ± 2 |
| 1.0 | 47 ± 5 | 35 ± 2 |
| 2.0 | 41 ± 5 | 35 ± 2 |

Milling the samples for longer milling times resulted in an appreciable broadening of the reflections and a reduction of the diffracted intensity while maintaining the general hexaferrite structure as shown in Fig. 4a for the sample with x = 2. The average crystallite size for this sample was reduced down to (27 nm) for the sample milled for 8 h, and to (21 nm) for that milled for 16 h. The variation of the crystallite size with milling time for x = 2 sample is shown in Fig. 4b. Similar behavior was also observed for the other samples with x = 0 and x = 1.

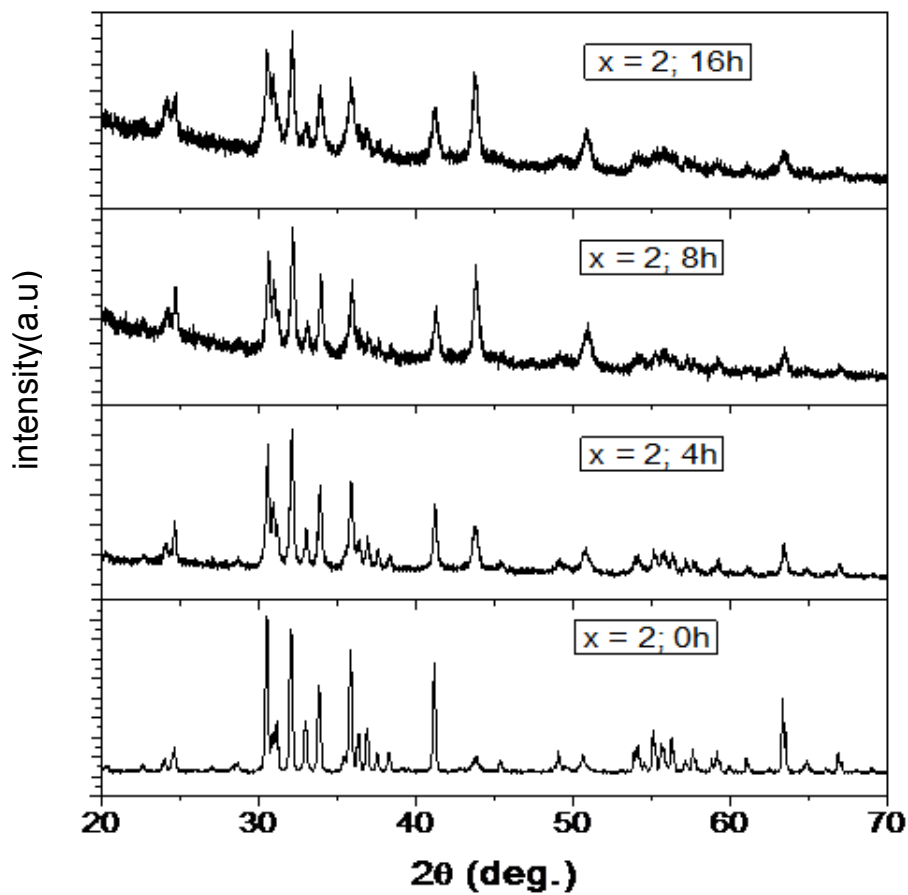

FIG. 4a. XRD pattern for $Ba_2 Zn_2Fe_{12}O_{22}$ with longer milling times for crystallite size analysis

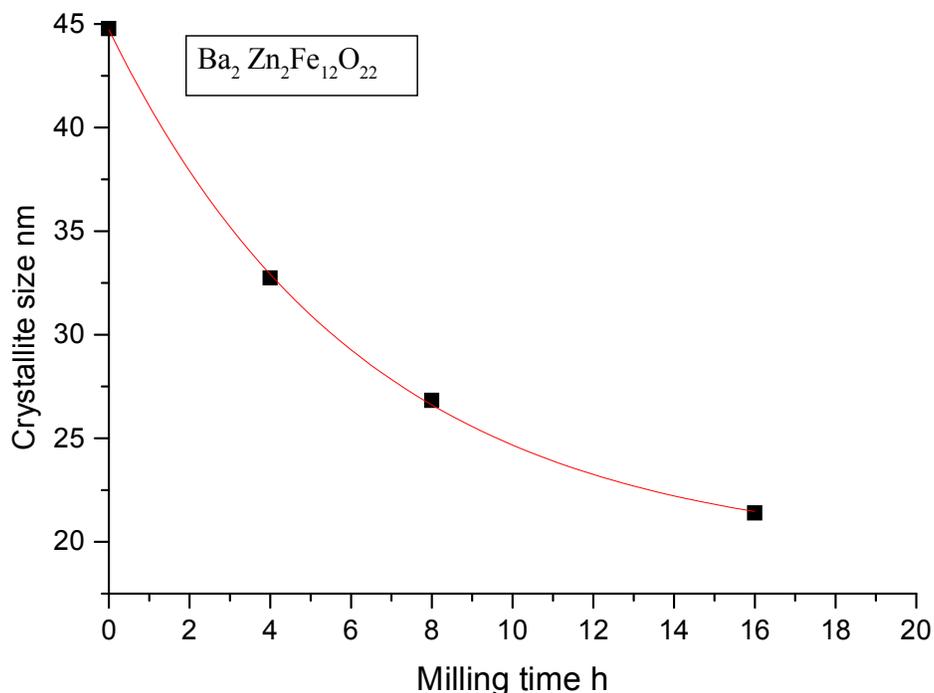

FIG. 4b. The variation of crystallite size for $Ba_2 Zn_2Fe_{12}O_{22}$ with milling time

**Scanning Electron Microscopy (SEM)**

SEM photos for the un-milled and milled ferrite samples (4 h milling time) are shown in Figs. 5, 6 and 7 for $x = 0$, 1 and 2, respectively. The images indicate that the milling process resulted in a change of the shape of the particles from platelet-like to a more or less spherical shape. In addition, the milled samples are composed of small particles which are about an order of magnitude smaller than the original particles. The particles in the un-milled samples have diameters ranging between 1 and 4 μm, while the samples milled for 4 h show that the majority of the particles have diameters in the range between 150 and 250 nm, with a small fraction of larger platelet-like particles, which indicates that the milling time was not sufficient for the production of a homogeneous powder. The physical particle size determined by SEM imaging is significantly larger than the crystallite size determined from the analysis of the XRD line profiles. This is indicative of the presence of crystal defects resulting in the formation of poly-crystalline particles.

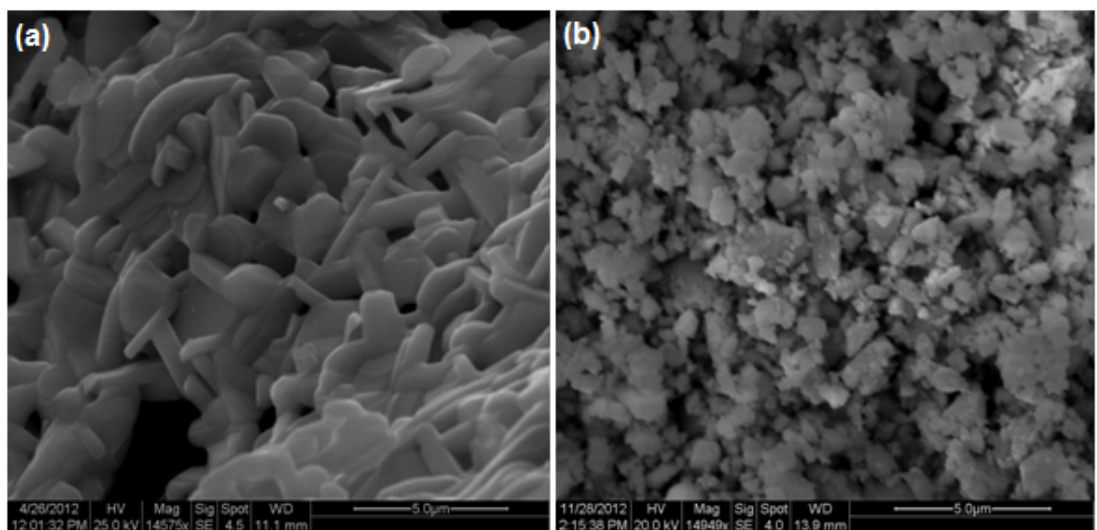

FIG. 5. SEM images for the sample with $x = 0$. (a) The un-milled sample, and (b) The sample milled for 4 h

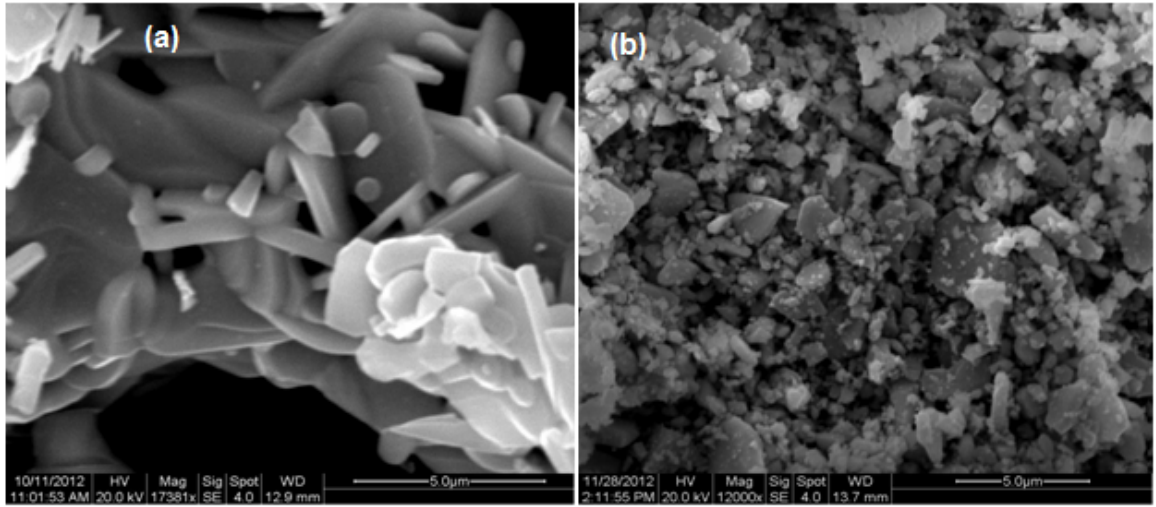

FIG. 6. SEM images for the sample with *x* = 1. (a) The un-milled sample, and (b) The sample milled for 4 h

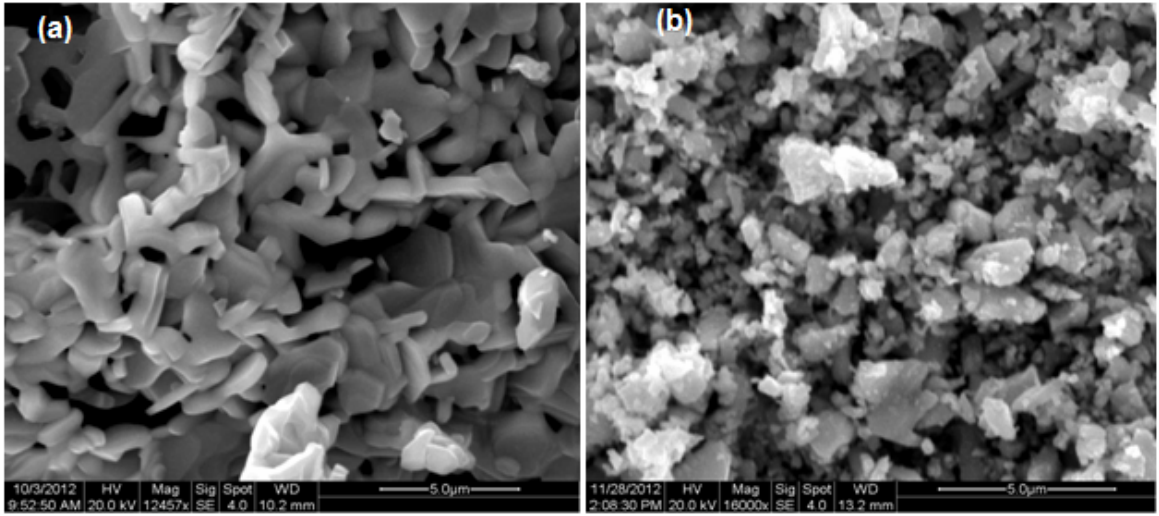

FIG. 7. SEM images for the sample with *x* = 2. (a) The un-milled sample, and (b) The sample milled for 4 h

**Mössbauer Spectroscopy**

Room temperature Mössbauer spectra of the samples of $Ba_2Zn_xCo_{2-x}Fe_{12}O_{22}$ ($x$ = 0, 1 and 2) with different milling times (T = 0, 0.5, 1, 2, 4 and 8 h) were recorded and then fitted using dedicated component fit routines. The spectra for the samples with $x$ = 0 with different milling times are shown in Fig. 8. The figure shows similar spectral structure with slightly different spectral characteristics for the sample milled for different times. This is consistent with the results of XRD measurements which indicate that the Y-type hexaferrite phase does not change at four hour milling and is still a pure phase. Mössbauer spectra for the samples milled up to 4 h were fitted with three sextet components. The high field component (I) is associated with $Fe^{3+}$ ions at the ($6c_{IV}^*$ + $3b_{VI}$) sites, the middle component (II) with the $18h_{VI}$ sites and the low field component (III) with the ($3a_{VI}$ + $6c_{VI}$ + $6c_{IV}$) sites in accordance with the assignment adopted in [17]. The spectrum of the sample milled for 8h was best fitted with the three above mentioned magnetic sextets and a broad, central paramagnetic component. This component could be an indication of the presence of small superparamagnetic particles in the powder, which were produced by virtue of the excessive grinding.

The values of the hyperfine parameters for the three-component fit are listed in Table 3 (T = 0 h). From the spectrum of the un-milled sample, the first component (I) with $B_{hf}$ = 49.0 T and a relative intensity of 20 % is associated with the ($6c_{IV}$* + $3b_{VI}$) sites as indicated above. The second component (II) with $B_{hf}$ = 46.4 T and a relative intensity of 45 % is associated with the $18h_{VI}$ sublattice, while the third component (III) with $B_{hf}$ = 42.9 T and a relative intensity of 35 % is attributed to the combination ($3a_{VI}$ + $6c_{VI}$ + $6c_{IV}$) sites.

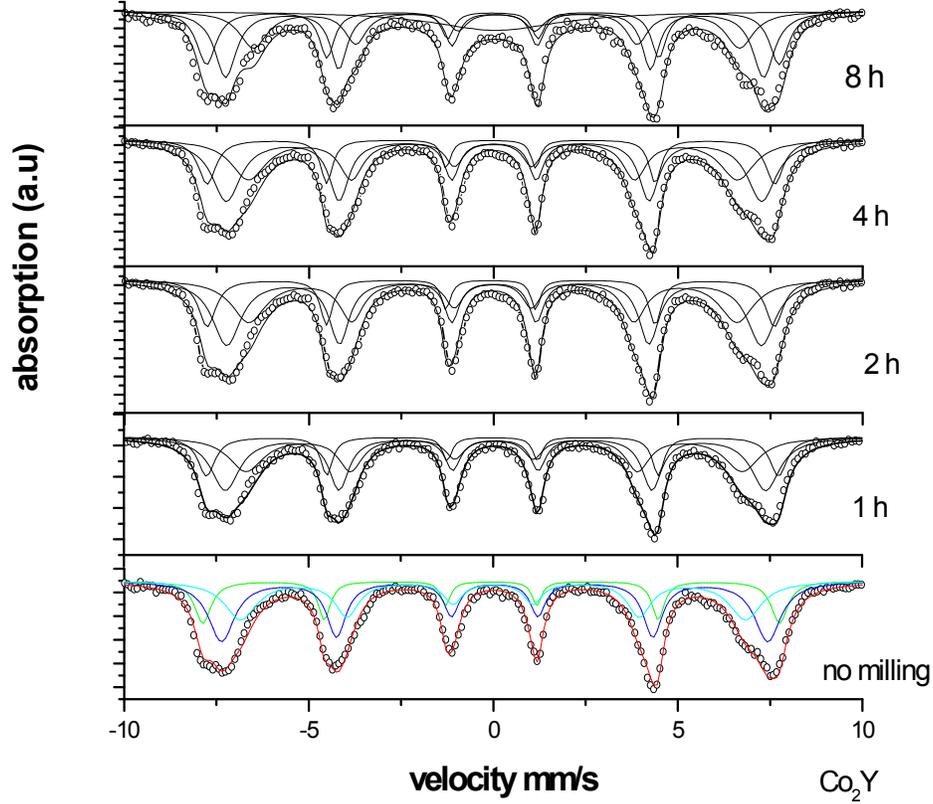

FIG. 8. Room temperature Mössbauer spectra of the sample $Ba_2Co_2Fe_{12}O_{22}$ sintered at temperature T = 1100° C and different milling times (T = 0, 1, 2, 4 and 8 h)

TABLE 3. Room temperature Mössbauer parameters of $Ba_2Co_2Fe_{12}O_{22}$ sintered at T = 1100° C, hyperfine field $B_{hf}$, center shift CS, relative intensity of the spectral component and line width (W) for inner lines

| Mössbauer Parameters | T = 0 h | T = 1 h | T = 2 h | T = 4 h | T = 8 h |
|---|---|---|---|---|---|
| $B_{hf}$ I(T) ± 0.5 T | 49.0 | 48.6 | 48.4 | 48.2 | 48.3 |
| $B_{hf}$ II(T) ± 0.5 T | 46.4 | 45.9 | 45.5 | 45.5 | 45.4 |
| $B_{hf}$ III(T) ± 0.5 T | 42.9 | 42.1 | 41.4 | 41.5 | 41.2 |
| QQ(mm/s) ± 0.04 mm/s | | | | | 0.06 |
| CS I (mm/s) ± 0.02 mm/s | 0.3 | 0.34 | 0.28 | 0.32 | 0.31 |
| CS II (mm/s) ± 0.02 mm/s | 0.39 | 0.42 | 0.36 | 0.31 | 0.38 |
| CS III (mm/s) ± 0.02 mm/s | 0.34 | 0.37 | 0.33 | 0.33 | 0.10 |
| CS QQ(mm/s) ± 0.02 mm/s | | | | | 0.41 |
| Relative Intensity I (%) | 20 | 20 | 20 | 20 | 25 |
| Relative Intensity II (%) | 45 | 45 | 45 | 45 | 35 |
| Relative Intensity III (%) | 35 | 35 | 35 | 35 | 25 |
| Relative Intensity QQ (%) | | | | | 15 |
| Width I (W) (mm/s) | 0.31 | 0.3 | 0.30 | 0.29 | 0.39 |
| Width II (W) (mm/s) | 0.50 | 0.50 | 0.48 | 0.49 | 0.46 |
| Width III (W) (mm/s) | 0.61 | 0.61 | 0.58 | 0.59 | 0.62 |
| Width QQ (W) (mm/s) | | | | | 4.74 |

For no milling time of x = 0, the maximum capacity of the ($6c_{IV}^*$ + $3b_{VI}$) sites for $Fe^{3+}$ ions is 3 ions per formula (25% of the 12 iron ions in a formula unit), and the maximum capacity of the $18h_{VI}$ site is 6 ions (50%), while the maximum capacity of the ($3a_{VI}$ + $6c_{VI}$ + $6c_{IV}$) sites is 5 ions (41.7%). The relative intensity of 45% of the middle component suggests that 5.4 $Fe^{3+}$ ions per formula occupy the $18h_{VI}$ sites and the relative intensity of 20% and 35% of the low and high field components are consistent with 2.4 $Fe^{3+}$ ions occupying the ($6c_{IV}^*$ + $3b_{VI}$) sites and 4.2 $Fe^{3+}$ ions occupying and the ($3a_{VI}$ + $6c_{VI}$ + $6c_{IV}$) sites, respectively.

It was previously reported that $Co^{2+}$ ions prefer to reside in face sharing octahedral sites in the T block ($6c_{VI}$ and $3b_{VI}$) [2]. However, it was reported by others that $Co^{2+}$ ions occupy $18h_{VI}$ and $6c_{IV}$ sites with marked preference for the $18h_{VI}$ octahedral sites [20]. Our results on the relative sub-spectral intensities of the three components indicate that the $Co^{2+}$ ions are distributed almost randomly over the sites corresponding to these components. The relatively higher $Co^{2+}$ ions for sites corresponding to the low field component could be interpreted as due to the preference of face-sharing sites due to the lower charge of this ion, which leads to lowering the electrostatic energy of the crystal structure.

Upon milling for 8 h, the relative intensities deviate appreciably from the above values and indicate that $Co^{2+}$ ions are almost randomly distributed over octahedral sites in the T block and at the S-T interface.

Mössbauer spectra at room temperature for the samples with x = 1.0 ($Ba_2ZnCoFe_{12}O_{22}$) milled for different milling times are shown in Fig. 9. The spectra for the samples milled up to 4h were fitted with three sextet components as in the previous case, with the high field component (I) associated with $Fe^{3+}$ ions at the ($6c_{IV}^*$ + $3b_{VI}$) sites, the middle component (II) with the $18h_{VI}$ site and the low field component (III) with the ($3a_{VI}$, $6c_{VI}$ and $6c_{IV}$) sites. The values of the hyperfine parameters for the samples are listed in Table 4. The appreciable drop in hyperfine field with Zn substitution is consistent with $Zn^{2+}$ ions occupying $6c_{IV}^*$, thus weakening the $3b_{VI}$ - $6c_{IV}^*$ and $18h_{VI}$ - $6c_{IV}^*$ antiferromagnetic interactions. The relative intensities for the un-milled and milled samples are the same as in the previous samples with x = 0. This is an indication that a fraction of $Zn^{2+}$ ions (with preference for tetrahedral coordination) occupy $6c_{IV}^*$ sites as an equal fraction of $Co^{2+}$ ions were removed from $3b_{VI}$ sites and replaced by $Fe^{3+}$ ions, maintaining the original relative intensity for the high field component. Similarly, a fraction of $Zn^{2+}$ ions occupy $6c_{IV}$ tetrahedral sites as an equal fraction of $Co^{2+}$ ions were removed from $6c_{VI}$ octahedral sites and replaced by $Fe^{3+}$ ions [20], maintaining the relative intensity of the low field component.

TABLE 4. Room temperature Mössbauer parameters of $Ba_2ZnCoFe_{12}O_{22}$ sintered at T = 1100° C, hyperfine field $B_{hf}$, center shift CS, relative intensity of the spectral component and line width (W) for inner lines

| Mössbauer Parameters | T = 0.0 h | T = 1h | T = 2 h | T = 4h | T = 8h |
|---|---|---|---|---|---|
| $B_{hf}$I(T) ± 0.5 T | 46.1 | 45.7 | 45.2 | 45.4 | 45.2 |
| $B_{hf}$II(T) ± 0.5 T | 42.2 | 41.4 | 41.2 | 41.3 | 41.0 |
| $B_{hf}$III(T) ± 0.5 T | 38.7 | 38.8 | 38.3 | 37.8 | 37.8 |
| QQ(mm/s) ± 0.04 mm/s | | | | | 0.08 |
| CS I (mm/s) ± 0.02 mm/s | 0.36 | 0.33 | 0.35 | 0.34 | 0.38 |
| CS II (mm/s) ± 0.02 mm/s | 0.4 | 0.24 | 0.33 | 0.33 | 0.36 |
| CS III (mm/s) ± 0.02 mm/s | 0.4 | 0.22 | 0.38 | 0.38 | 0.41 |
| CS QQ(mm/s) ± 0.02 mm/s | | | | | 0.42 |
| Relative Intensity I (%) | 20 | 20 | 20 | 20 | 25 |
| Relative Intensity II (%) | 45 | 45 | 45 | 45 | 35 |
| Relative Intensity III (%) | 35 | 35 | 35 | 35 | 25 |
| Relative Intensity QQ (%) | | | | | 15 |
| Width I (W) (mm/s) | 0.33 | 0.31 | 0.35 | 0.35 | 0.41 |
| Width II (W) (mm/s) | 0.51 | 0.43 | 0.50 | 0.50 | 0.46 |
| Width III (W) (mm/s) | 0.61 | 0.48 | 0.76 | 0.77 | 0.60 |
| Width QQ (W) (mm/s) | | | | | 3.06 |

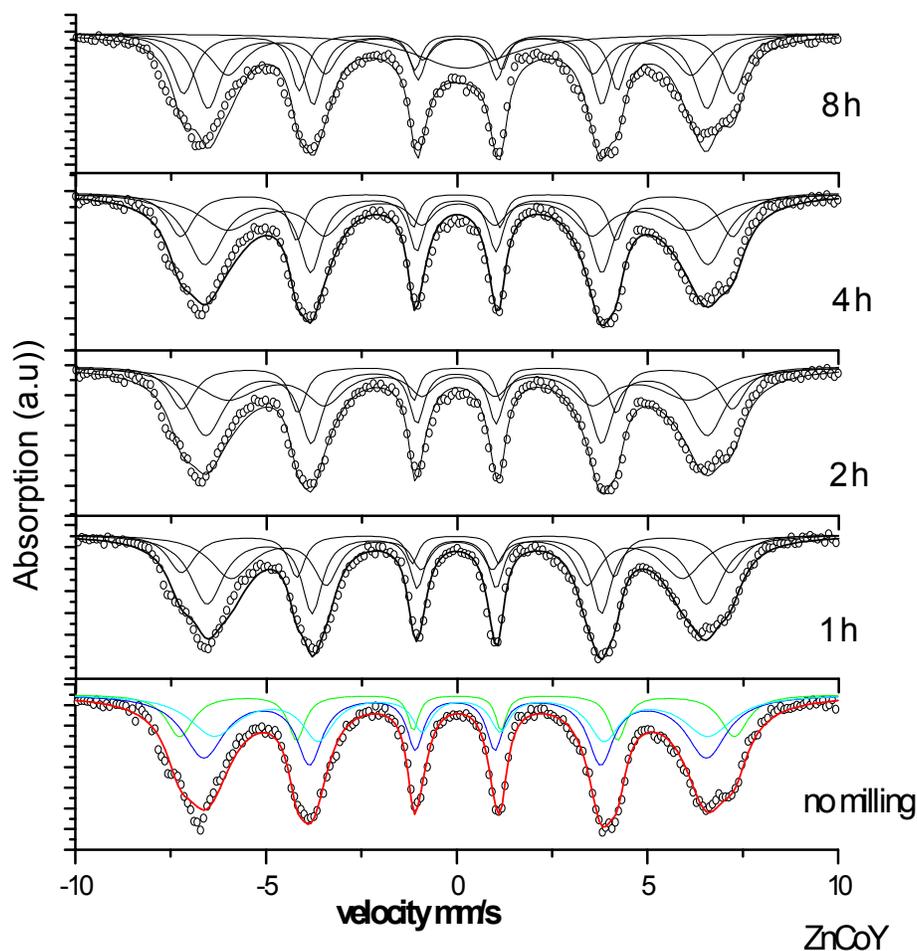

FIG. 9. Room temperature Mössbauer spectra of the sample $Ba_2ZnCoFe_{12}O_{22}$ sintered at temperature T = 1100°C and different milling times (T = 0, 1, 2, 4 and 8 h)

Mössbauer spectra at room temperature for the sample with x = 2 ($Ba_2Zn_2Fe_{12}O_{22}$) milled at different milling times (T = 0, 1, 2, 4 and 8 h) are shown in Fig. 9. The spectra were fitted with three magnetic sextets (except for the 8h milled sample which was fitted with three magnetic sextets and one quadrupole doublet). The corresponding hyperfine parameters are listed in Table 5. The relative intensity of 50% of the middle component indicates a complete filling of the $18h_{VI}$ sites by $Fe^{3+}$ ions. However, the relative intensities (23% and 27%) of the other two components suggest that 0.24 $Zn^{2+}$ ions in this sample occupy the $6c_{IV}$ sites and 1.76 ions occupy the $6c_{IV}^*$ sites resulting in the observed large drop of the intensity of the corresponding component down to 27%.

The values of $B_{hf}$ for the three components (I, II and III) are plotted *versus* milling time for the samples with three $Zn^{2+}$ concentrations (x = 0, 1 and 2) as shown in Figs. 11-13.

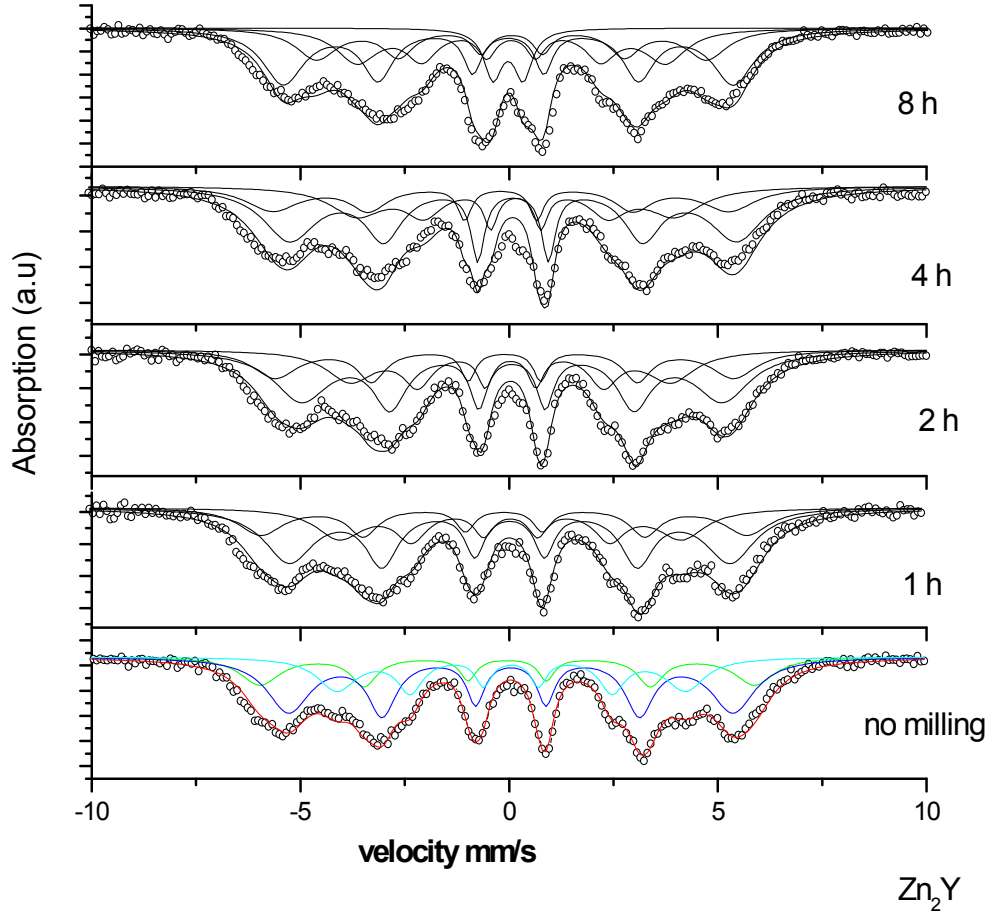

FIG. 10: Room temperature Mössbauer spectra of the sample $Ba_2Zn_2Fe_{12}O_{22}$ sintered at temperature $T = 1100^\circ$ C and different milling times (T = 0, 1, 2, 4 and 8 h)

Table 5: Room temperature Mössbauer parameters of $Ba_2Zn_2Fe_{12}O_{22}$ sintered at $T = 1100\ ^\circ C$, hyperfine field $B_{hf}$, center shift CS, relative intensity of the spectral component and line width (W) for inner lines

| Mössbauer Parameters | T = 0.0 h | T = 1h | T = 2 h | T = 4h | T = 8h |
|---|---|---|---|---|---|
| $B_{hf}$ I (T) ± 0.5 T | 37.1 | 36.5 | 34.4 | 34.3 | 33.8 |
| $B_{hf}$ II (T) ± 0.5 T | 33.4 | 33.1 | 31.6 | 30.6 | 29.3 |
| $B_{hf}$ III (T) ± 0.5 T | 26.2 | 25.7 | 24.1 | 23.9 | 23.2 |
| QQ (mm/s) ± 0.04 mm/s | | | | | 0.7 |
| CS I (mm/s) ± 0.02 mm/s | 0.31 | 0.23 | 0.24 | 0.25 | 0.33 |
| CS II (mm/s) ± 0.02 mm/s | 0.39 | 0.36 | 0.42 | 0.35 | 0.41 |
| CS III (mm/s) ± 0.02 mm/s | 0.4 | 0.38 | 0.39 | 0.41 | 0.4 |
| CS QQ (mm/s) ± 0.02 mm/s | | | | | 0.01 |
| Relative Intensity I (%) | 23 | 23 | 23 | 23 | 24 |
| Relative Intensity II (%) | 50 | 50 | 50 | 50 | 38 |
| Relative Intensity III (%) | 27 | 27 | 27 | 27 | 22 |
| Relative Intensity (QQ) (%) | | | | | 16 |
| Width I (W) (mm/s) | 0.45 | 0.53 | 0.42 | 0.39 | 0.51 |
| Width II (W) (mm/s) | 0.51 | 0.57 | 0.52 | 0.39 | 0.51 |
| Width III (W) (mm/s) | 0.46 | 0.55 | 0.44 | 0.39 | 0.51 |
| Width QQ (W) (mm/s) | | | | | 0.51 |

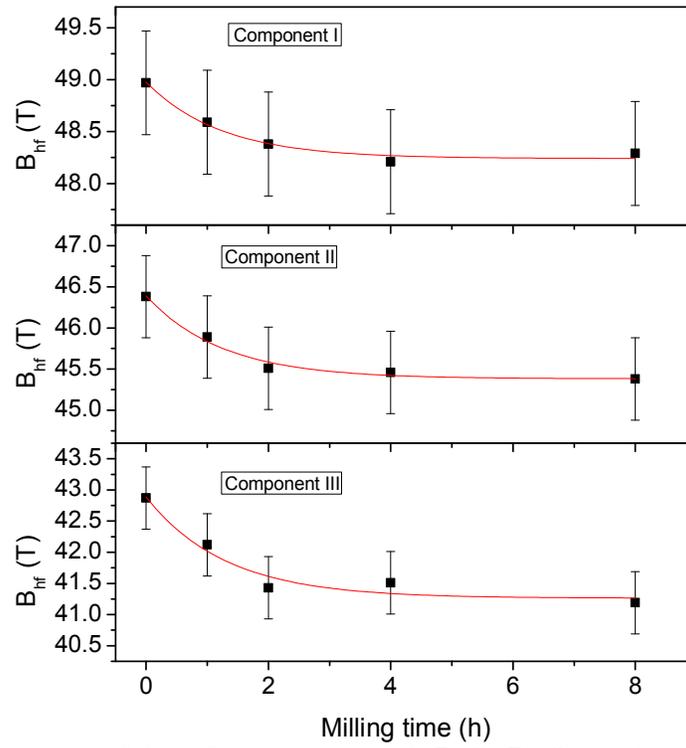

FIG. 11. The distribution of hyperfine parameters of $Ba_2Co_2Fe_{12}O_{22}$ with different milling times (T = 0, 1, 2, 4 and 8 h)

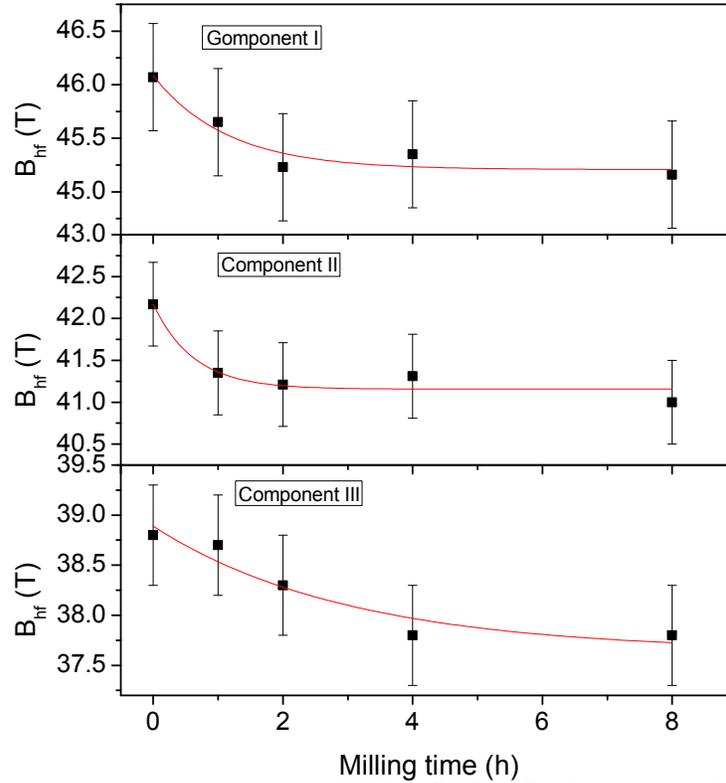

FIG.12. The distribution of hyperfine parameters of $Ba_2ZnCoFe_{12}O_{22}$ with different milling times (T = 0, 1, 2, 4 and 8 h)

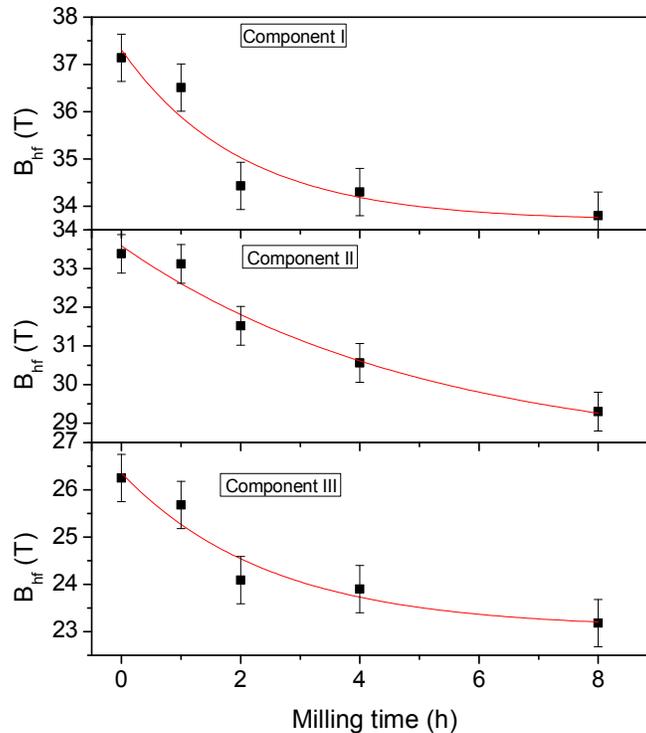

FIG. 13. The distribution of hyperfine parameters of $Ba_2Zn_2Fe_{12}O_{22}$ with different milling times (T = 0, 1, 2, 4 and 8 h)

The values of the hyperfine parameters for (x= 0, 1 and 2) are decreasing exponentially with increasing milling time. The spectrum broadening is due to the presence of distributions of hyperfine fields which is partially attributed to the statistical distribution of neighbors around the iron sites in the lattice. This effect is a consequence of the mechanical milling which results in crystal defects and could change the distribution of $Fe^{3+}$ ions over the different sites of the hexaferrite lattice. Further, the increasing in milling time leads to the increase of the ratio of surface atoms to volume atoms. This results in the reduction of the hyperfine field of the magnetic components corresponding to surface atoms, resulting in asymmetric broadened lines. The new component represented with the quadrupole doublet in the spectrum of the 8 h milled sample is associated with small superparamgnetic grains resulting from the relatively long ball-milling time, as confirmed by XRD and SEM results.

## Conclusions

The system $Ba_2Zn_xCo_{2-x}Fe_{12}O_{22}$ with (x = 0, 1 and 2) was prepared by sol-gel method. Milling in different parts of time (T = 0, 1, 2, 4 and 8h), XRD patterns indicate that the information and the analyses show that the system $Ba_2Zn_xCo_{2-x}Fe_{12}O_{22}$ with (x = 0, 1 and 2) is a Y-type hexaferrite and it maintains the pure Y-type with milling up to 8 h.

Mössbauer spectra at room temperature for all the samples $Ba_2Zn_xCo_{2-x}Fe_{12}O_{22}$ with x= 0, 1 and 2 sintered at T = 1100º C show a gradual change in the relative intensities and a drop in the hyperfine field as $Zn^{2+}$ ions replacing $Co^{2+}$ ions increase. The variations in relative intensities are consistent with $Co^{2+}$ ions occupying octahedral sites and $Zn^{2+}$ ions occupying tetrahedral sites.

The spectra of $Ba_2Zn_xCo_{2-x}Fe_{12}O_{22}$ with (x = 0, 1 and 2) were fitted by 3 sextet components up to 4 h milling. After milling for 8 h, Mössbauer spectra were best fitted with three magnetic sextets and a quadrupole component; which indicates the presence of superparamagnetic particles with largely reduced particle size in these samples. The $B_{hf}$ values decrease gradually with milling time, which is a consequence of the decrease in particle sizes with increasing the milling time.

## Acknowledgement

This work was supported by a generous grant from the Scientific Research Fund (SRF) in Jordan.